\PassOptionsToPackage{table}{xcolor}
\documentclass[sigconf]{acmart}

\renewcommand\footnotetextcopyrightpermission[1]{} 
\usepackage{xcolor}
\usepackage{booktabs} 
\usepackage{subfig}
\usepackage{graphicx}
\usepackage{hyperref}

\setcopyright{rightsretained}

\makeatletter
\def\@copyrightspace{\relax}
\makeatother

\begin{document}
\title[``Breaking'' Disasters]{``Breaking'' Disasters: Predicting and Characterizing the Global News Value of Natural and Man-made Disasters}


\author{Armineh Nourbakhsh, Quanzhi Li, Xiaomo Liu, Sameena Shah}
\affiliation{%
  \institution{Research \& Development}
  \institution{Thomson Reuters}
  \city{New York} 
  \state{NY} 
}
\email{{armineh.nourbakhsh,quanzhi.li,xiaomo.liu,sameena.shah}@tr.com}





\begin{abstract}
Due to their often unexpected nature, natural and man-made disasters are difficult to monitor and detect for journalists and disaster management response teams. Journalists are increasingly relying on signals from social media to detect such stories in their early stage of development. Twitter, which features a vast network of local news outlets, is a major source of early signal for disaster detection. Journalists who work for global desks often follow these sources via Twitter's lists, but have to comb through thousands of small-scale or low-impact stories to find events that may be globally relevant. These are events that have a large scope, high impact, or potential geo-political relevance. We propose a model for automatically identifying events from local news sources that may break on a global scale within the next 24 hours. The results are promising and can be used in a predictive setting to help journalists manage their sources more effectively, or in a descriptive manner to analyze media coverage of disasters. Through the feature evaluation process, we also address the question: ``what makes a disaster event newsworthy on a global scale?'' As part of our data collection process, we have created a list of local sources of disaster/accident news on Twitter, which we have made publicly available.

\end{abstract}

%
%



\maketitle

\section{Introduction}\label{intro}
Early reports of unexpected events are known as \textit{tips} among journalists. These are often propagated on local news media, social media, or through word of mouth in the early stages of an event's development. These tips are vital to journalists since they help them detect and report unexpected events in a timely manner. 

A topic that is particularly of interest is disasters and accidents, such as natural disasters, terror attacks, wars and military conflict, social unrest, or man-made accidents. Due to their geo-dependency, such events are often reported on a local scale before global media captures them. A big part of the daily routine of journalists who monitor these events is to sift through small-scale or low-impact stories and find ones that are relevant no a national or global scale. Characterizing the potential news value of a disaster can help us detect, contextualize and report these events automatically. In addition, understanding the aspects of a story that make it newsworthy can help us identify and analyze potential media bias \cite{kwak:2014}. This study proposes a model that predicts whether a disaster or accident that has been reported via local sources on social media will be reported by global news media within the next 24 hours.

There are several aspects along which a natural or man-made disaster can be characterized as low-profile or high-profile. Below we have listed a few possible characteristics. These aspects will later be used in the modeling and classification of high-profile vs. low-profile stories. 
\begin{itemize}
    \item An obvious parameter in determining the newsworthiness of an event is its \textbf{topic}. Unexpected disasters such as earthquakes may be more newsworthy than expected ones such as snowstorms. Geo-politically-sensitive incidents such as terror attacks may draw more attention than other criminal activity, and so on. 
    \item A disaster can happen at a small or large \textbf{scale}. A large wildfire has higher news value than a house fire. 
    \item The \textbf{impact} (or potential impact) of a disaster may influence the way media report it. This can be \textit{human impact} (such as the number of dead, injured, missing, displaced, or affected people), \textit{physical impact} (such as collapsed infrastructure), or \textit{financial impact} (such as disrupted business, economic cost, or monetary loss).
    \item The \textbf{location} of an event may be related to the attention it receives from global media. 
    \item Some events are more common in certain locations. For instance, Japan is more prone to earthquakes than France. Shootings are more common in the U.S. than Australia. The \textbf{rarity} of an event can have an impact on whether it is picked up by global media.
\end{itemize}

In the following sections, we review relevant literature in the domain of event detection and contextualization from social media, lay out our methodology, describe our features, discuss and evaluate our models. We show that a small but strong set of features can have a robust and promising performance, and discuss the relevance of particular features to the classification task. As part of our data collection process, we have curated a list of users who report disasters and accidents on a regular basis, which we have made publicly available. 

\section{Related Work}
Twitter's significance as a source of breaking news has been demonstrated in previous studies \cite{osborne:2014}. Research on early event detection from social media has largely followed an unfiltered approach, i.e. it relies on an open stream of tweets, possibly filtered by keywords or geo-tags. First Story Detection (FSD) methods attempt to cluster streaming messages in real-time, while Burst Detection methods wait for a story to show a spike in volume \cite{atefeh:2015}. These approaches offer quick detection of stories, and can be used as a proxy to determine the newsworthiness of an event. For instance, the rate of growth of clusters of tweets that discuss an event can be used to predict its newsworthiness. However, due to the time-sensitive nature of most high-profile disasters, it is important to flag newsworthy stories as soon as possible. In such cases, waiting for clusters to bypass a certain threshold of growth may not be an option. An alternative is to rely entirely on the immediate content of the story to determine its news value.   

Content-based studies of newsworthiness have been applied to both long-form and short-form text \cite{upadhyay:2016,freitas:2016,piotrkowicz:2016}. They use endogenous and exogenous features to predict the news value of a given document. Freitas and Ji \cite{freitas:2016} apply a classification model to detecting newsworthy tweets. Since their model is not focused on a particular topic, they use the compositional features of tweets (e.g. slang usage and sentimental tone) as well as presence of named entities (e.g. geopolitical entities or company names). Their analysis is not concerned with the scope of a story's newsworthiness (e.g. local vs. global).

Kwak and An \cite{kwak:2014} analyze the front-page coverage of stories from news media around the world. Using \textit{GDELT}'s global index of local, national and global news outlets\footnote{http://www.gdeltproject.org/about.html\#creation}, they inspect various contextual, topical, and sociopolitical factors and their potential relation to whether the story is covered on a global scale. One of the most significant factors in their study is previous coverage by global news agencies such as \textit{Reuters} and \textit{Associated Press}.

Our study mirrors \cite{kwak:2014}, in that it attempts to predict whether a given story will be reported by global news agencies. Rather than using the \textit{GDELT} dataset, we use Twitter, because the former only provides data at the end of each day, by which time it is too late for global monitoring and editorial desks to make a decision on whether or not to report a story. Another reason for using Twitter is the fact that it offers data from sources that are not necessarily captured by the media indexed in the \textit{GDLET} project. These include blog posts, fire and police departments, and smaller news outlets that operate at the local level. 

In addition to offering a larger set of sources, Twitter also provides events in their ``pre-breaking'' stage, i.e. before they are picked up and reported by mainstream media. Local agencies such as fire and police departments, security offices, local authorities, local radio and TV stations, are invaluable sources of breaking news \cite{liu2016reuters}. The language of early reports about an event is often different from headlines published about it. Table \ref{tab:early} shows a few examples of early Twitter reports about unexpected incidents and contrasts them with news headlines. Compared to the headlines, the early reports follow a different tone, format, and are more ambiguous on details such as casualties.

\begin{table*}
\centering
\caption{Four examples of early disaster reporting on Twitter by local sources, versus the first headline published to Reuters' internal wire. From top to bottom the early reporters are a local journalist, a local news outlet, a local fire department, and a local authority.}
\label{tab:early}
\begin{tabular}{l|l}
\textbf{Early tweet} & \textbf{Reuters News Headline}                                              \\ \hline \hline
    \raisebox{-.5\height}{\includegraphics[width=2.5in]{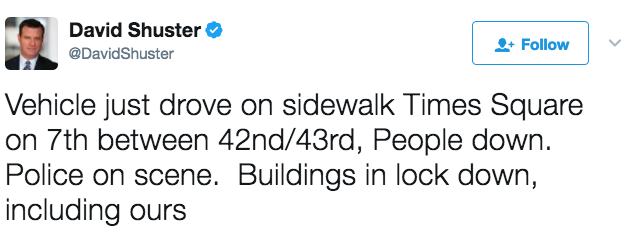}}                 & \begin{tabular}[c]{@{}l@{}}``Speeding vehicle strikes pedestrians in\\New York City's Times Square''\end{tabular}    \\ \hline
    \raisebox{-.5\height}{\includegraphics[width=2.5in]{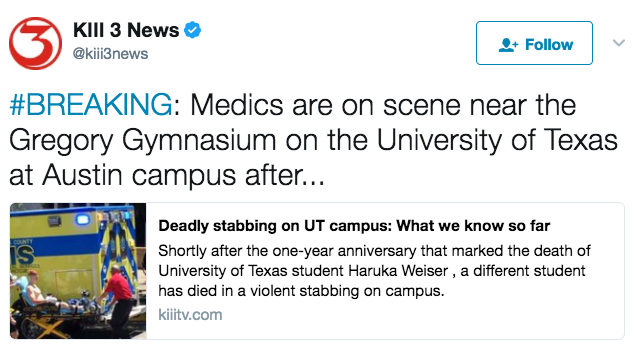}}                 & \begin{tabular}[c]{@{}l@{}}``Multiple people stabbed at University of\\Texas in Austin -police''\end{tabular}        \\ \hline
    \raisebox{-.5\height}{\includegraphics[width=2.5in]{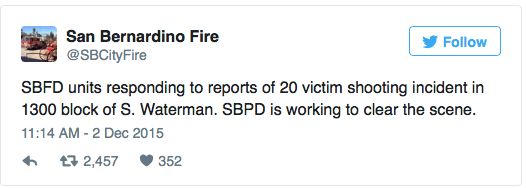}}                 & \begin{tabular}[c]{@{}l@{}}``Reports of 20 victims wounded in shooting\\in San Bernardino, California''\end{tabular} \\ \hline
    \raisebox{-.5\height}{\includegraphics[width=2.5in]{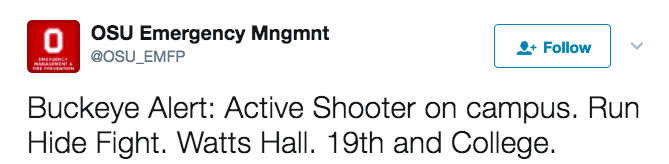}}                 & \begin{tabular}[c]{@{}l@{}}``Active shooter on campus at Ohio State\\University''\end{tabular}                      
\end{tabular}
\end{table*}


In this study, we present a method for predicting whether an early Twitter report of a natural or man-made disaster is globally relevant. By focusing on endogenous features that are relevant to the immediate context of the story, we try to minimize the effect of exogenous factors that make the model dependent on external knowledge bases. We also refrain from using any user-dependent features to keep the model source-agnostic, however the coverage, quality and reliability of sources can sway the performance of the model. We have discussed user credibility and event verification in a separate publication \cite{liu2015rumor}.

\section{Problem Statement}
Journalists who monitor news on Twitter often use two main sources. One possible tool is keyword-based streams. Journalists can create Boolean queries based on their topics of interest (e.g. ``bomb OR blast OR explosion'') and set up streams of tweets that match the query. They can use tools such as TweetDeck\footnote{https://tweetdeck.twitter.com} and HootSuite\footnote{https://hootsuite.com} for easier set up and refinement of such streams. 

Another major source are Twitter lists. Twitter allows users to organize other users into lists. For instance, a journalist who is interested in the war in Yemen may create a list that consists of reporters who are covering the war, another list for humanitarian agencies involved, and so on. Lists are an important tool for news-gathering, since they often consist of vetted and reliable sources. Compared to keyword-based streams, user-based lists offer less noise, more relevant content, and are more reliable. Most of them are professional accounts that only report on certain newsworthy topics. Nevertheless, they still suffer from the problem of information overload (See Table \ref{tab:users}). Some accounts (like earthquake monitors) send automated notifications of seismic activity on a regular basis. Other monitoring accounts (such as ``@Breaking911'') post reports of accidents and disasters from around the world. These may be relevant to local news outlets, but for journalists who work for national or global outlets, it is difficult to sift through the information in a timely and efficient manner, and find events that will break on a global scale. 


Given a disaster event in its early stage of development, the goal of this study is to predict whether the event will be reported by any of five global media outlets, namely \textit{Reuters}, \textit{Associated Press}, \textit{Agence France-Presse}, \textit{CNN}, and \textit{BBC} within the next 24 hours. A disaster event is defined as one falling under one of the following topics as defined by the TRBC classification system\footnote{https://financial.thomsonreuters.com/en/products/data-analytics/market-data/indices/trbc-indices.html}: \textit{floods}, \textit{earthquakes and other seismic activity}, \textit{severe weather conditions}, \textit{fires and explosions}, \textit{acts of terrorism}, \textit{wars and military conflicts}, \textit{violence and crime}.

\section{Data}
To curate the data required for this study, we used \textit{Reuters News Tracer}, a real-time news detection and verification engine \cite{Stray:2016}. \textit{Tracer} uses a FSD algorithm to spot breaking stories in real time from Twitter \cite{li:2017}. Each event is represented as a cluster of tweets that discuss that story.

The purpose of this study is not to analyze \textit{Tracer}'s output, but to simulate lists that journalists use to monitor disasters. As a result, we did not directly use the clusters generated by the tool. Instead, we used them to curate a large list of users who regularly report on disaster events around the world. The list was curated automatically and enriched with metadata about each user's location, affiliation, the type of event they report on, and other standard metrics such as followers and friends. 

\subsection{Collecting locally-focused users}\label{sec:local}
To collect the list, we first exported 140,723 clusters generated in December of 2016, which were labeled as disaster events in \textit{Tracer}. We selected the earliest tweet from each cluster and added the user of that tweet to a preliminary list. Then, for each user, we identified the location on their profile and obtained lat/lon information for each location using the \textit{Nominatim} geo-tagger\footnote{https://nominatim.openstreetmap.org}.

Next, we removed the following users from the list:
\begin{itemize}
    \item Users with more than 1 million followers: These are often national news outlets (such as CNN or BBC) that should not be included because 1) they are not limited to disaster reporting, and 2) even when they report a disaster, it is already a ``mature'' event which has been captured and reported at the local level. 
    \item Users with no location assignment in their profile: This was necessary, since we were only interested in locally-focused accounts.
    \item Users whose profile location did not match their content: Locally-focused accounts often post about their location of interest. For instance, the New York City Fire Department often posts about incidents in New York City. To distinguish between these accounts and those that weren't locally focused, we sampled 50 random tweets from each profile. We identified the location of each tweet using the geo-parser provided by \textit{OpenCalais} \footnote{http://www.opencalais.com}. 
    
    \textit{Nominatim} allows guided queries, where a toponym is submitted for geo-tagging, but its lookup is only allowed within a certain region\footnote{http://wiki.openstreetmap.org/wiki/Nominatim\#Parameters\_2}. We submitted each tweet location to \textit{Nominatim}, using the profile location as ``anchor.'' If \textit{Nominatim} was able to find the tweet location within the profile location, the search was considered a hit. Otherwise it was considered a miss. Accounts that had a hit-to-miss ratio of 0.5 or higher were considered locally-focused.
\end{itemize}

\subsection{Collecting topically-focused users}
The above steps were helpful in identifying locally-focused accounts, but they were likely to remove Twitter monitors from the dataset (such as earthquake monitors or accounts like ``@Breaking911''). These accounts can have a large following, and are usually not locally focused. In order to add them back into the dataset, we ran the discarded accounts through a processor that determined whether they were topically focused. We did this by looking up each account in \textit{Tracer}. \textit{Tracer} tags each cluster by one of 11 news topics. Accounts that were overwhelmingly tagged by ``Law/Crime'' or ``Crisis/War/Disaster'' would be added back into the dataset. To define what ``overwhelmingly'' means in this context, we calculated a \textit{tf.idf} score for each account \cite{manning:2009}, where the account represents a \textit{term} and each topic represents a \textit{document}. Accounts whose \textit{tf.idf} score for the two given topics was in the top 20 percentile were considered topically-focused, and added back to the dataset.

This left us with 10,607 individual users. 

\subsection{Characterizing the user accounts}\label{dict}
After curating the list of users, we added the following metadata to each user:

\subsubsection{Informativeness} 
We assigned a score to each user based on their informativeness. Informativeness was measured by the number of disaster/accident stories that each account had participated in, per 100 tweets. 
\subsubsection{Type of account}\label{trbc} 
Each account was classified into one of 8 categories: \textit{fire (or emergency services)}, \textit{police and traffic reports}, \textit{local authorities} (such as office of the mayor), \textit{local news} (including TV/radio stations, print media and news websites), \textit{local journalists} (people affiliated with local news media), \textit{earthquake monitors}, \textit{severe weather monitors}, and \textit{disaster monitors}. 

To predict each user's category, we used topic codes from Reuters News headlines. Reuters archives news stories in a system that enriches each story with metadata tags, including Thomson Reuters Business Classification (TRBC) codes. These include codes for \textit{fire/explosions}, \textit{earthquakes/seismic activity}, \textit{violence/crime}, \textit{terrorism/insurgency}, \textit{war/military conflict}, \textit{floods}, \textit{severe weather events}, and \textit{disaster/accidents}.

We randomly sampled 1,000 headlines for each TRBC, and 1,000 tweets from each account (or all of the account's tweets, if fewer than 1,000 tweets had been posted), creating a global dictionary for headlines and tweets \footnote{Sampling from the TRBC codes had to be modified to address a problem with overlap. The codes follow a hierarchy and are not exclusive. Each story that is tagged with a lower-level tag, gets tagged with all of its ancestors as well. For instance since \textit{disaster/accidents} is an ancestor of \textit{fire/explosions}, any story that's tagged as a fire automatically gets a disaster tag as well. To remove the overlap, for higher-level tags like the disaster tag, we only included headlines that did not include any of its descendants.}. Next, we coded each headline into a \textit{tf.idf} vector of its terms, where each TRBC code was treated as a document. We represented each TRBC code by a centroid vector. We followed a similar logic for each user, sampling tweets from each account and representing the account by a centroid. We calculated the pairwise cosine distance between each user centroid and each TRBC centroid, assigning the user to its nearest TRBC code. For instance, accounts closest to the \textit{fire/explosion} code were labeled as fire departments. 

We randomly sampled 20 accounts from each category and evaluated the labels.The analysis correctly labeled \textit{fire departments}, \textit{police departments}, \textit{earthquake monitors}, and \textit{severe weather monitors}, with 87\% accuracy. However, \textit{local news outlets}, \textit{local journalists}, \textit{local authorities}, and \textit{disaster monitors} were grouped together (all relating to the \textit{disaster/accident} code). To distinguish them, we first identified journalists by their use of personal pronouns or occupational nouns in their profile description.

Next, we located global disaster monitors by determining if they were locally-focused (as described in section \ref{sec:local}). Finally, we separated local authorities from local news media by simply looking up the words ``news,'' ``newsdesk'', ``newsroom,'' ``headlines,'' ``press,'' ``coverage,'' ``channel'', ``station,'' ``TV,'' ``television,'' and ``radio'' in their name or profile description. 

We have made the list of accounts available as a Kaggle dataset\footnote{https://www.kaggle.com/arminehn/disasteraccident-sources}. Table \ref{tab:users} lists some statistics about the users. Weather and earthquake monitors are the most informative categories, especially the latter which offers the largest value of informativeness over its small size. Overall, the accounts produce close to one million tweets per day, which is impossible to monitor manually. 

\begin{figure*}
\includegraphics[scale=0.8]{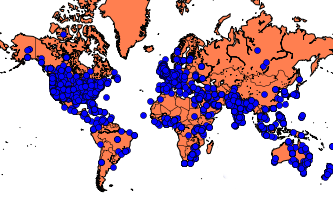}
\caption{Global projection of users in the disaster dataset.}\label{fig:map}
\end{figure*}

Figure \ref{fig:map} shows a global mapping of each user's location. Despite an expected bias towards North America and Europe, the dataset has a relatively large global coverage. Outside Europe and English-speaking countries, the coverage is biased towards regions under geo-political crisis, or prone to natural disasters. This is likely due to the fact that the dataset was curated from a collection of disaster-related tweets.

\begin{table*}
  \caption{Statistics about the user dataset.}
  \label{tab:users}
  \begin{tabular}{lcccll}
    \toprule
    User category&Count&\begin{tabular}{@{}c@{}}Avg.\\infomativeness\end{tabular}&\begin{tabular}{@{}c@{}}Avg. tweets\\per day\end{tabular}&Example account&Example tweet\\
    \midrule
    local news & 3,909& 2.58&449,315&@abs13houston&\begin{tabular}{@{}l@{}}``\#BREAKING Third suspect arrested\\after deadly shooting near Alvin''\\\end{tabular}\\ \hline
    local journalist & 3,792& 1.46&286,967&@NatashaFatah&\begin{tabular}{@{}l@{}}``\#BREAKING Black smoke pouring\\from beneath Scotia Plaza in heart of Toronto's\\business district''\\\end{tabular}\\\hline
    fire/emergency & 702& 1.79&34,102&@cityofwsfire&\begin{tabular}{@{}l@{}}``Residential house fire 3910 Tangle Ln. \#wsfire .80''\\\end{tabular}\\\hline
    police/traffic & 557 & 1.87&48,368&@MetroPoliceUK&\begin{tabular}{@{}l@{}}``Incident in \#Westminster: Please report\\anything suspicious to the Anti-Terrorist\\Hotline 0800 789 321.''\\\end{tabular}\\\hline
    local authority & 204& 1.40&17,571&@HarrisCountyDAO&\begin{tabular}{@{}l@{}}``HCDAO news advisory regarding a hate-crime\\charge that was filed today:Prosecutors\\Allege Hate Crime in Attack on African American''\\\end{tabular}\\\hline
    disaster monitor & 512& 1.96&50,382&@TerrorEvents&\begin{tabular}{@{}l@{}}``\#Syria \#Mayadeen - US-led strikes kill 35\\civilians in east Syrian town, held by \#IS: monitor''\\\end{tabular}\\\hline
    quake monitor & 25& 2.80&2,165&@USGSBigQuakes&\begin{tabular}{@{}l@{}}``Prelim M5.8 earthquake off the\\coast of Jalisco, Mexico May-20 06:02 UTC''\\\end{tabular}\\\hline
    weather monitor & 906& 3.22&64,657&@severewarn&\begin{tabular}{@{}l@{}}``Severe t-storm for Eastern Amite County\\and West Pike County for winds of \\60 mph and quarter sized hail.\\Move indoors away from windows.''\end{tabular}\\\hline
    All & 10,607&2.0&953,527&\cellcolor{gray!25}&\cellcolor{gray!25}\\
  \bottomrule
\end{tabular}
\end{table*}

\subsection{Finding disasters from the user feed}
We set up a feed of the given accounts, collecting their tweets in the month of April, 2017. We removed tweets that did not include any of the words in the news dictionary collected in section \ref{dict}. The resulting 65,417 tweets were split into two collections. 20,000 tweets were set aside to guide the feature-generation process. This dataset will be referred to as the ``booster dataset'' in the following section. The remaining 45,417 tweets were used for training, validation, and testing.  

\section{Methodology}
After collecting the tweets, the first step was to find proper labels for them. Due to the large size of the corpus, we implemented a method for noisy labeling.

As reported in \cite{liu2016reuters}, given a pool of potentially newsworthy tweets, the cosine similarity between the vectorized representation of tweets and news headlines is a good indicator of a match. The study found that a threshold of 0.5 for similarity can recall 68.9\% of matches with 90\% precision (headlines were extracted from AP, Reuters and CNN). We followed a similar approach and matched each tweet in the dataset against headlines published by Reuters, AP, AFP, CNN, and BBC in a 24-hour period after the tweet was posted. Tweets that matched a headline posted before them, but no headline posted after them, were considered to be ``tardy'' and remained unmatched.

As will be described in the following sections, many of the features used in the classification are taxonomy-based. In order to avoid a conflation between the labeling process and the feature extraction process, prior to matching the tweets with the headlines, we flagged any taxonomy token that appeared in the tweet or the headline. The tokens were not removed, but masked with the name of the feature they represented (e.g. scope-related words were flagged with \textit{scope\_quake\_magnitude}, \textit{scope\_scale\_adj}, etc.). A random evaluation of 100 matched tweets and 100 unmatched tweets showed agreement with \cite{liu2016reuters} (P = 86.7\%, R = 78.8\%).

To improve recall, we used a message-linking process, i.e. we ran a cosine similarity analysis between unmatched tweets and matched ones. Any unmatched tweet that had a similarity of 0.5 or higher to a matched tweet from later the same day was added to the matched set (the threshold was reduced to 0.3 or for a matched tweet from the same user). This reduced the number of unmatched tweets by 23\%. The propagation process could be repeated iteratively, but we only performed it once to avoid any long-tail errors. 

The remaining dataset still had a class imbalance problem (only about 3\% of data was matched), so we under-sampled the unmatched messages to create a 1-to-10 ratio between matched and unmatched tweets. This reduced the size of the training set to 11,990. 


\subsection{Features}
Section \ref{intro} laid out five main dimensions along which an event's news value can be characterized. We automatically extracted features related to each dimension, as is described below:

\subsubsection{Topic}\label{top} Section \ref{trbc} discussed the TRBC codes used to tag each news headline in the Reuters archives. Following the approach described in that section, we used the same set of codes to label each tweet's topic. 
\subsubsection{Scope} 
We used seven separate indicators to detect the scope of an event:
    \begin{itemize}
        \item A lexicon of 21 adjectives indicatives of scale, severity, and magnitude. These were extracted from relevant entries in Roget's Thesaurus\footnote{These consisted of 8 adjectives from section I:III:1:31 (``greatness''), 3 adjectives from section I:I:2:3 (``substantiality''), 7 adjectives from section I:VIII:3:173 (``violence''), and 3 adjectives from section III:III:1:361 (``killing'').}.
        \item A regular expression that detects expressions of multiple-alarm fires (e.g. ``3-alarm fire reported,'' or ``requesting a 2nd alarm''). These are phrases that are commonly used by fire departments in USA and Canada to express the scope of a fire. 
        \item A taxonomy of common causes of accidental fires and explosions, such as ``trash fire,'' ``gas leak,'' or ``lightning.'' These were collected by running the booster dataset against the list of users and identifying 150 user accounts that belonged to fire departments. Among the tweets posted by those accounts, we identified 460 tweets that included one of the expressions ``caused by,'' ``identified as,'' ``determined as,'' or ``confirmed as.'' We extracted the noun phrases following those expressions and manually filtered expressions that were not relevant. This resulted in 15 terms. The purpose of this taxonomy was to help the classifier distinguish between small fires/explosions and large ones through their common causes.
        \item A magnitude extractor designed specifically for earthquakes (including Mercalli\footnote{https://earthquake.usgs.gov/learn/topics/mercalli.php}, Richter\footnote{https://earthquake.usgs.gov/learn/topics/measure.php}, EMS\footnote{http://media.gfz-potsdam.de/gfz/sec26/resources/documents/PDF/EMS-98\_short\_form\_English\_PDF.pdf}, Shindo (JMA)\footnote{http://www.jma.go.jp/jma/en/Activities/inttable.html}, and CSIS\footnote{https://en.wikipedia.org/wiki/China\_seismic\_intensity\_scale} scales).
        \item A size extractor designed to detect the size of wildfires, expressed in acres, square miles, square kilometers, or radii.
        \item A regular expression that detects the size of multi-vehicle crashes such as ``2-car crash,'' or ``2 commercial trucks \& one vehicle.''
        \item A scale-extractor for severe weather conditions such as tornadoes and storms. The extractor detected expressions of Enhanced Fujita scale\footnote{http://www.spc.noaa.gov/faq/tornado/ef-scale.html}, TORRO scale \footnote{http://www.torro.org.uk/tscale.php}, and Beaufort scale \footnote{http://www.stormfax.com/beaufort.htm}. In case of hailstorms, it also locates expressions of size or diameter\footnote{http://www.spc.noaa.gov/misc/tables/hailsize.htm}. 
    \end{itemize}
\subsubsection{Impact}
Impact is often expressed explicitly, as a number or figure. To detect these expressions, we first ran a numeric-phrase extractor on each tweet. The extractor detected numbers expressed numerically (e.g. ``12''), alphabetically (e.g. ``twelve''), or implicitly (e.g. ``a dozen''). Soft expressions such as ``scores of,'' or ``several'' were detected, as well as expressions of larger numbers such as ``thousands,'' ``lakh,'' or ``crore''. 

Some numeric expressions in disaster tweets are parts of a physical address or a date/timestamp. We trained a simple linear classifier to to distinguish between dates, addresses, expressions of human impact (such as casualties and displacements) and expressions of financial impact (such as damages and losses). Using TweetNLP \cite{Owoputi13} to extract noun phrases from the tweets, we discarded all phrases that had been covered by scope indicators. Among the remaining phrases, we labeled 2,000, and trained a SGD classifier on them. The labeling was done by two researchers (kappa=0.981). The features used were:
\begin{itemize}
    \item A Boolean indicator expressing whether the numeric phrase includes tokens that are mixtures of digits and letters (this reduces the likelihood that it is an expression of human impact). 
    \item Presence of currency symbols next to the number (e.g. ``\$120'').  
    \item Presence of monetary symbols next to the number (e.g. ``120MM'').
    \item Presence of a timestamp symbol within the number (e.g. ``-'' or ``:'').
    \item Presence of timezone or period indicators within the number (e.g. ``EDT'' or ``AM'').
    \item Presence of terms related to human impact such as death toll, or number of casualties, injured, wounded, hospitalized, missing, displaced, or evacuated people. To collect these expressions, we ran the booster dataset through the numeric phrase detector and collected all of the noun phrases that included numeric expressions. For each phrase, we removed the numeric part and created a frequency map of remaining tokens. Then, two researchers inspected the terms in the top five percentiles and collected tokens related to death and injury. This allowed us to go beyond common expressions and include terms such as ``lynched,'' ``martyred,'' ``drowned,'' or ``buried alive.'' 
    \item Presence of terms related to physical addresses such as ``bldg,'' ``rd,'' etc. These were collected from LipPostal's taxonomy of street addresses\footnote{https://github.com/openvenues/libpostal/tree/master/resources/dictionaries/en}.
    \item Vector representation of each tweet, where the vector was composed of \textit{tf.idf} presentation of each term in the tweet. In calculating the \textit{tf.idf}, each category (address, human impact, financial impact) was treated as a document. Each vector was reduced to three elements representing the tweet's maximum \textit{tf.idf} value for one of the categories.
\end{itemize}

The classifier was trained using 5-fold cross-validation. The resulting performance was good (F1 = 89.54, P = 92.57, R = 86.71) so we did not explore other models. The simplicity of the features and the effectiveness of the linear classifier indicate that the classification (at least in the context of disaster tweets) is a simple task that may be converted into a rule-based algorithm.  

Another type of impact that may be relevant to the profile of an event is physical impact. This type of impact can indicate whether a vital infrastructure has been affected (e.g. the collapse of a bridge) or an industry has been disrupted (e.g. explosion at a refinery). Physical impact is directly linked to the site of the disaster. The site of the disaster is different from its location. Terms such as ``apartment,'' ``barn,'' ``restaurant,'' ``trailer,'' ``airport,'' ``office,'' ``highway,'' ``bridge,'' ``refinery,'' ``school,'' ``church,'' etc. indicate the site.  

TRBC includes topic codes for business and infrastructure sites such as \textit{oil/gas refineries}, \textit{pipelines}, \textit{aircrafts}, \textit{tanker freight}, etc. We used the TRBC definition of each code as a guided taxonomy. After tokenizing the descriptions and tagging their parts-of-speech using the NLTK library package\footnote{http://www.nltk.org/book/ch05.html}, we used the nouns as features denoting physical sites. 

\subsubsection{Location} We used \textit{OpenCalais} to tag each tweet for its location. Each location was encoded by it latitude and longitude, as well as its name and its two-letter country code\footnote{https://en.wikipedia.org/wiki/ISO\_3166-1\_alpha-2}, resulting in four different features. This allowed the classifier to match locations on more than one dimension.

If no location was found in the tweet but it came from a locally-focused user, then the user's location was substituted. If the user was not locally-focused, then all location-related features were set to \textit{nil}.
\subsubsection{Rarity} So far all of the features discussed were endogenous features (i.e. depended only on the immediate content of the tweet). However, some exogenous context, like the commonality of a certain type of disaster at a certain location, can impact its news value. For instance, a shootout is a very rare occurrence in Belgium, but a more common one in the United States. Or an earthquake is far more common in Japan than Qatar. Modeling the potential effect of rarity can help the classifier decide the news value of an event.

The rarity of an event can change over time, for instance a country engaged in war is prone to events related to bombs and military conflict, but that may not be the case before or after the war. To model this effect, we used a 3 month period (Jan-Mar 2017) as the background to the training period. Rarity was defined as:

\begin{equation}
Rarity_e = \frac{|T^{3m}_{<l_e,s_e>}|}{|T^{3m}_{<l_e>}|} + \lambda \frac{|T^{3m}_{<L_e,s_e>}|}{|T^{3m}_{<L_e>}|}
\end{equation}

Where $<l_e,s_e>$ is a pair representing an event $e$, $l_e$ is the location of the event expressed by its lat/lon coordinates, $s_e$ is the topic of the event, $L_e$ is the location of the event expressed by its country code, $T^{3m}_{<x,y>}$ is the set of tweets posted in the previous 3 months that are tagged by $x$ and $y$\footnote{We did not want rarity to encode a potential reporting bias, so instead of using Reuters headlines, we used tweets to model an event's rarity.}, and $LR_e$ is the location-dependent rarity of $e$. $\lambda$ is a parameter that discounts the effect of country-level reporting compared to the location-specific reporting by trying to ``guess'' how effectively the country name be used as a proxy for the specific location's name:

\begin{equation}
\lambda = \frac{|T^{3m}_{<L_e, l_e>}|}{|T^{3m}_{<L_e>}|}
\end{equation}

Rarity can have a location-independent dimension as well. Prior to the terrorist attack in Nice on July 14, 2016 where a truck was deliberately driven into a large crowd, accidents involving vehicles were not always deemed globally newsworthy. However after that attack, and when similar attacks took place in the United States and Germany, global journalists have been more conscious of similar incidents. This type of rarity is incident-specific and requires additional language modeling that can connect the main agents of an incident to previously observed events. We have not implemented this in the current version of this study.


\subsection{Experiments and Evaluation}
\subsubsection{Performance} 
We cross-validated a linear SVM classifier on 10-folds with 80\%-20\% splits. Table \ref{tab:eval} shows the performance of the model using various feature sets. The results show that indicators of impact and scope offer a big improvement to recall, while rarity and location are less effective in that regard. The features have a strong synthetic effect when combined. Contrary to expectations, despite the small number of features and the class imbalance problem, the classifier performs strongly on recall.  

\begin{table}
\centering
\caption{Performance of a linear SVM model using different feature sets.}
\label{tab:eval}
\begin{tabular}{l|c|c|c}
\textbf{Model} & \textbf{P}     & \textbf{R}     & \textbf{F}      \\ \hline
\begin{tabular}[c]{@{}l@{}}baseline\\(\textit{tweet\_vector+topic})\end{tabular}          & 63.01          & 51.46          & 56.65                                                                                          \\ \hline
\begin{tabular}[c]{@{}l@{}}\textit{tweet\_vector+topic}\\\textit{+scope+impact}\end{tabular}         & 73.58          & 75.48 & 74.52                                                                          \\ \hline
\begin{tabular}[c]{@{}l@{}}\textit{tweet\_vector+topic}\\\textit{+rarity+location}\end{tabular}       & 72.30          & 60.99          & 66.17                                                                                          \\ \hline
\begin{tabular}[c]{@{}l@{}}All features\end{tabular}       & \textbf{85.68}          & \textbf{80.93}          & \textbf{83.24}                                                                                          \\
\end{tabular}
\end{table}

\begin{figure}
\includegraphics[scale=0.45]{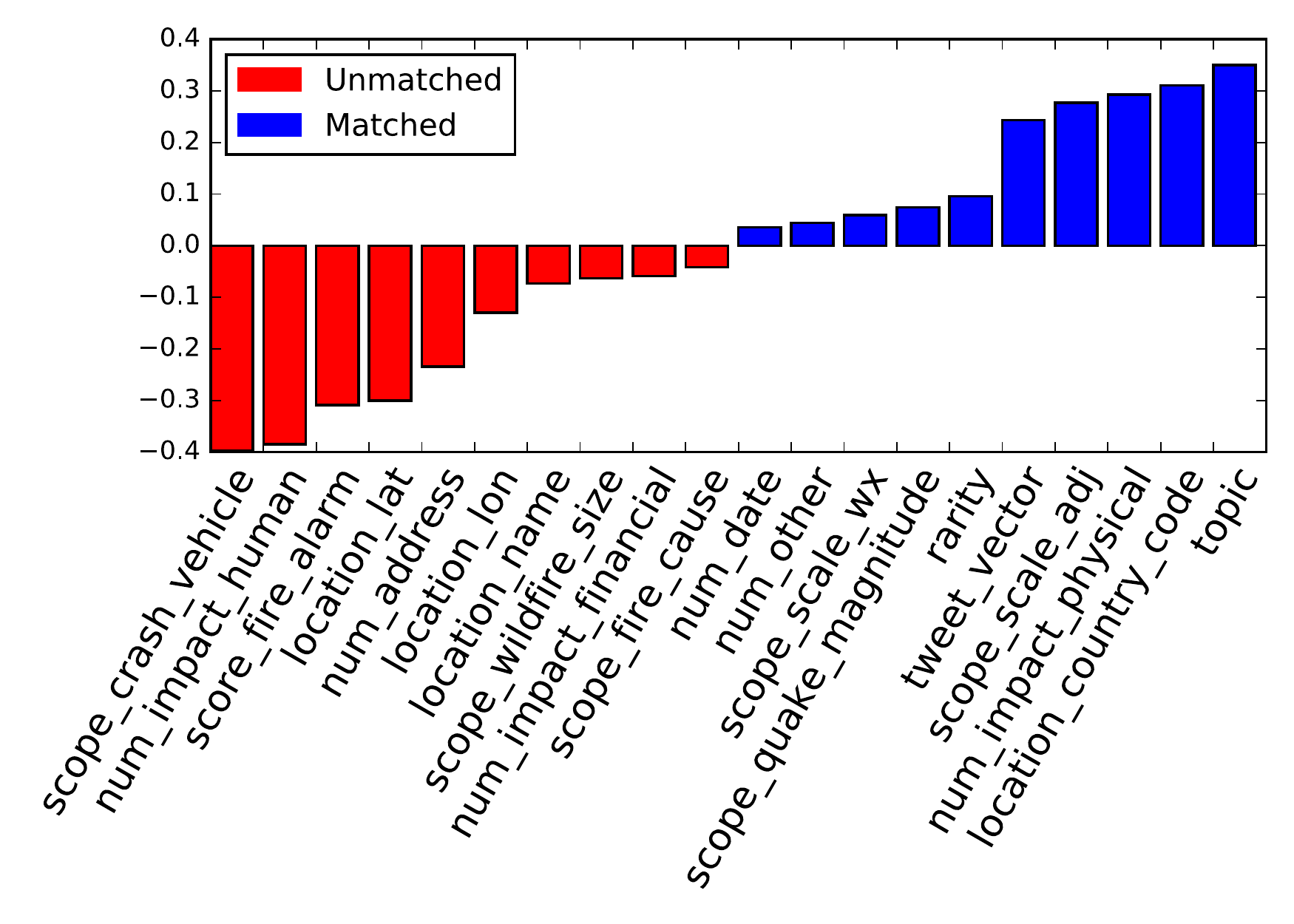}
\caption{Feature groups and their weights in the SVM model.}\label{fig:features}
\end{figure}

Figure \ref{fig:features} shows the weights of feature sets in the model. As expected, topic is the biggest contributor to the matched class, followed by the country code. Contrary to expectation, human impact seems to be a better predictor of a non-match. This may have been caused by the fact that many stories do not have an expression of human impact until they have matured. Tweets often report events in their early stage of development when there is not much information available. Once the event has matured and the human impact confirmed, news headlines begin reporting the casualties, but by this point any tweet reporting the event will be ``tardy'' and thus unmatched.

Sometimes the expressions of impact are present but vague, e.g. in Table \ref{tab:early}, the tweet reporting the UT Austin campus stabbing describes it as ``deadly,''\footnote{Adjectives such as ``deadly'' are covered in the scope\_scale\_adj taxonomy, which may be why this feature has a high ranking in the matched class.} and in the tweet reporting the Times Square incident, the human impact is expressed as ``people down.'' More sophisticated language modeling and NLP techniques are required to capture these expressions.

Rarity did not play a big part in the Matched class. This may have been caused by geo-political factors, e.g. global media frequently reports on wars and military conflicts, so a bomb explosion in a city involved in a war, despite not being a rare event, is still reported globally \cite{kwak:2014}.


\subsubsection{Timeliness}
\begin{figure}
\includegraphics[scale=0.34]{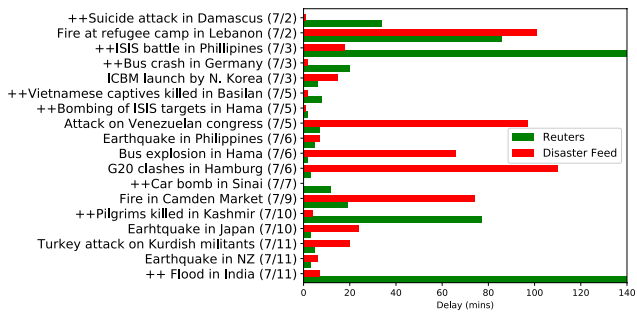}
\caption{Delay of the disaster feed versus Reuters News in reporting events. The time of each event has been approximated by looking up the earliest tweet posted about it on Twitter. Events where the disaster feed was ahead of Reuters have been highlighted by ++.}\label{fig:timeliness}
\end{figure}

We tested the system's timeliness by running it on the live feed of disaster users between 07/01/2017 and 07/11/2017. At the end of the period, we downloaded the feed of news alerts from Reuters News and filtered it down to disaster/accident events using their TRBC topic codes. This resulted in 18 events. Figure \ref{fig:timeliness} shows the individual events and the delay of Reuters versus the disaster feed in reporting them. On average, the disaster feed was 27 minutes ahead of Reuters, and beat Reuters 44\% of times.

\section{Conclusion and Future Work}
In this study, we presented a method for predicting the global news value of a natural or man-made disaster. Our study indicates that a small set of features related to the scope, impact, and location of the event drives its global news value. 

This study was only focused on predicting breaking disasters from a feed of local authorities and local news media. Eyewitnesses who are at the scene of a disaster are another major source of news \cite{liu2016reuters}. There is a growing body of research around identifying eyewitness accounts from social media \cite{morstatter:2014, doggett:2016, fang:2016} and we hope to incorporate some of that research into future work. 

A media outlet, by way of its ownership, affiliation, location, or other factors, may show interest in certain news items. We deliberately avoided any analysis of the impact of previous media interest for two reasons: Media interest may be the result of inherent reporting bias which can seep into the model's parameters. Excluding previous media reports also helps keep the model independent from its prediction target, so that its predictions can be used in three ways: A) To help journalists spot potentially newsworthy stories, or B) To model a ``snapshot'' of one news outlet's preferences, and show potential deviation from that snapshot over time, or C) To train separate models for separate news outlets and compare the models analytically and descriptively. 

Another matter that is not addressed in this study is verification. Many stories do not break on a global scale because they are deemed unreliable, or are debunked by authoritative sources. For example, on 03/18/2017 an explosion in a residential area in the Saint Gilles municipality of Brussels drew attention due to its suspicious nature\footnote{https://twitter.com/AncaGurzu/status/843116074451308545}. However, it was soon confirmed as a gas explosion and remained unreported at the global scale. We have discussed the problem of automatic verification of news stories propagated on Twitter in real time in another publication \cite{liu2015rumor}.


\bibliographystyle{ACM-Reference-Format}
\bibliography{reference} 

\end{document}